\documentclass[iop,revtex4]{emulateapj}
\usepackage{lineno}

\begin{document}

\renewcommand{\topfraction}{1.0}
\renewcommand{\bottomfraction}{1.0}
\renewcommand{\textfraction}{0.0}

\title{Spectroscopic subsystems in nearby wide binaries}

\author{Andrei Tokovinin}
\affil{Cerro Tololo Inter-American Observatory, Casilla 603, La Serena, Chile}
\email{atokovinin@ctio.noao.edu}

\begin{abstract}
Radial velocity (RV) monitoring of solar-type visual binaries has been
conducted  at the  CTIO/SMARTS 1.5-m  telescope to  study short-period
systems.  Data reduction is described,  mean and individual RVs of 163
observed objects are given.  New spectroscopic binaries are discovered
or suspected in 17 objects, for  some of them orbital periods could be
determined.  Subsystems  are efficiently  detected  even  in a  single
observation by  double lines and/or  by the RV difference  between the
components  of  visual  binaries.   The potential  of  this  detection
technique  is  quantified  by  simulation  and  used  for  statistical
assessment  of 96 wide  binaries within  67\,pc. It  is found  that 43
binaries  contain  at  least  one  subsystem  and  the  occurrence  of
subsystems  is  equally  probable   in  either  primary  or  secondary
components.  The  frequency of subsystems and their  periods match the
simple prescription proposed  by the author (2014, AJ,  147, 87).  The
remaining 53  simple wide binaries with a  median projected separation
of 1300\,AU have  the distribution of the RV  difference between their
components  that  is  not  compatible with  the  thermal  eccentricity
distribution  $f(e)=2e$ but  rather matches  the  uniform eccentricity
distribution.
\end{abstract}


\maketitle

\section{Introduction}
\label{sec:intro}

Binary and multiple systems  are ubiquitous products of star formation
\citep{DK13}.   The  gas  condensing  into  stars  contains  excessive
angular  momentum that  is  transported outwards  during collapse  and
accretion   and  stored   partially   in  the   orbital  momentum   of
binaries. While the basic physics  is known, no current theory is able
to  predict  the  multiplicity   fraction  and  the  distributions  of
separations,  mass  ratios,  or eccentricities,  although  large-scale
numerical  simulations  show  promising  agreement  with  observations
\citep{Bate2012}.  Formation of binary stars is related to the stellar
mass  function  and sets  the  scene  for  subsequent evolution  (mass
transfer, blue stragglers, X-ray  binaries, and type I supernovae). On
the other hand, the gas left around newly born stars as a material for
planet formation is inherited from  the epoch of mass accretion and is
related  to multiplicity.  The  common origin  of stars,  planets, and
binaries should therefore be studied jointly.

Solid observational data on  binary statistics are difficult to
get because of various selection  effects, and this is exacerbated for
stellar  hierarchies  containing  three   or  more  stars.   Yet,  
hierarchical multiple systems   give
additional  insights into  star  formation from  their period  ratios,
relative orbit orientation, and  structure.  This work complements the
study of  hierarchical multiplicity of  solar-type stars in  the 67-pc
volume  \citep[][hereafter FG67a  and FG67b]{FG67a,FG67b}.   The large
size of the FG-67 sample defined in FG67a  (4867 stars) and modest and well-controlled observational
selection allowed  reliable estimates of the  fraction of hierarchical
systems, 13\%, and of their statistics.  One unexpected result was the
 large fraction of  subsystems in the secondary components of
wide binaries and the emerging correlation between subsystems, so that
4\% of targets turned out to  be 2+2 quadruples (two close binaries on
a  wide orbit  around  each  other). However,  this  result relied  on
extrapolation  from   the  small  number   of  well-studied  secondary
components.   While the  main targets  of  the FG-67  sample were  the
subject  of  spectroscopic and  imaging  observational campaigns,  the
secondaries  were  generally  neglected.   Recent imaging  studies  of
secondary  components   \citep{RAO,SAM}  have  largely   remedied  the
situation,  but {\it  close} secondary  subsystems accessible  only to
spectroscopy  still remain  to be  discovered.  Exploring  these close
subsystems is the main motivation of the present work.

I report  here RV monitoring of  solar-type stars in  the FG-67 sample
that are either known to  be spectroscopic binaries but lack the orbit
or  belong  to  wide  visual  binaries.  Most  targets  have  southern
declinations.  The main goal  is characterization of  subsystems in
the   secondary  components  and   determination  of   their  periods,
complementing  previous efforts  in this  respect. Extensive
previous RV  data were furnished by the  Geneva-Copenhagen Survey, GCS
\citep{N04}, which covered mostly the main targets.  As a continuation
of this  campaign, \citet{Halb2012} monitored  the RVs of  wide pairs,
discovering several  subsystems.  \citet{TS02} reported  RV monitoring
of  both wide  and close  (unresolved  on the  slit) visual  binaries,
mostly  on the northern  sky. A  similar, but  much smaller  survey of
southern multiples is  published by \citet{LCO}.  \citet{Desidera2006}
measured  precise  RVs  of  wide  binaries in  search  of  stellar  or
planetary  subsystems.  These  survey-type works  are  complemented by
studies  of individual objects  published by  various authors
\citet[e.g.][]{Griffin2001}. 
Although this paper is  devoted to wide binaries, RVs of
all  observed  stars  are   given.   Presentation  of  newly
determined  orbits is  deferred to  a future  paper, as  their lengthy
discussion would distract from  the main topic, while binaries needing
more data are still being observed.

The  instruments,  data  reduction,  and  raw  results  are covered in
Section~\ref{sec:obs}.   Then  in  Section~\ref{sec:simul} I  evaluate
through simulation the proposed  method of detecting subsystems from a
single  observation.  Section~\ref{sec:widebin} reports  on subsystems
in a volume-limited sample of wide binaries.  In Section~\ref{sec:ecc}
the RV difference  in wide binaries without subsystems  is analyzed to
constrain   their   eccentricities.    The   work   is   summarized   in
Section~\ref{sec:sum}.

\section{Observations and results}
\label{sec:obs}

\subsection{Observing campaigns and instruments}

The observations reported  in this paper were obtained  with the 1.5-m
telescope  located at  the Cerro  Tololo Interamerican  Observatory in
Chile    and   operated    by    the   SMARTS    Consortium.\footnote{
  \url{http://www.astro.yale.edu/smarts/}}  The   observing  time  was
allocated  through  NOAO  (programs 10B-0022,  14B-0009,  and  15A-0055).   The
observations were made by the telescope operators in service mode.

The goal of  the 2010 campaign was to  determine unknown spectroscopic
orbits of close solar-type binaries  and to observe components of wide
binaries that lacked  RV  coverage. AT least three observations
of each target were planned in the 20 allocated nights.  However, less
time was actually scheduled, so most
targets were observed  just once, and only a  few were pointed several
times.  This data set is  therefore too small for addressing the goals
of the  original proposal. Nevertheless,  one  orbit was  determined and
several subsystems were discovered.

In  2010,  the  spectra  were  taken with  the  Fiber  Echelle  (FECH)
instrument.  It was the  de-commissioned echelle spectrometer from the
Blanco 4-m telescope moved to the Coud\'e room of the 1.5-m telescope
and fed by the fiber. In this program the image of the fiber projected
into the spectrograph was not masked by the slit and its width defined
the spectral resolution of $R=44\,000$.  The detector was a SIT CCD of
2014$\times$2048  pixels  format   with  the  Arcon  controller.   The
calibration spectrum  of the thorium-argon  lamp fed through  the same
fiber was taken before or after each target.

In 2014 and 2015, observations  were made with the CHIRON spectrograph
\citep{CHIRON}  that replaced FECH  in 2011.   Stars brighter  than $V
\sim  9.5$  mag were  observed  in the  slicer  mode  with a  spectral
resolution of  $R=90\,000$, while for  fainter targets the  fiber mode
with $R=28\,000$ was used.  When both components of a wide binary were
observed, the  same mode  was chosen.  The  thorium-argon calibrations
were recorded for each target.

The observing campaigns of 2014 and 2015 targeted secondary components
of  wide  binaries  in   a  ``snapshot  survey''  designed  to  detect
subsystems  from just  one observation  by RV  difference  between the
components.   I selected  binaries  from the  FG-67  sample south  of
$+20^\circ$  declination with  separations no  less than  8\arcsec ~to
ensure that  the components  are well resolved  on the  entrance fiber
aperture.  Only the  secondaries with $V$ magnitudes between  8 and 11
 were  chosen to shorten  the exposure time, maximizing  the sample
coverage.  Secondaries  with prior RV  coverage were removed  from the
program;  however, they are  included in  the statistical  analysis in
Section~\ref{sec:widebin}.  Primary  components were observed  as well
if they lacked prior RV data comparable in quality to the GCS (two or
three observations with a time base of at least a year).  Also, monitoring of
several  known  solar-type spectroscopic  binaries  was continued  for
computing their orbits.

\subsection{Data reduction}

The reduced and wavelength-calibrated  spectra delivered by the CHIRON
pipeline were retrieved from the SMARTS center at the Yale University.
The availability of this service  has greatly enhanced this program by
allowing  rapid analysis  of the  RVs and  flexible scheduling  of new
observations when needed.

\begin{figure}
\plotone{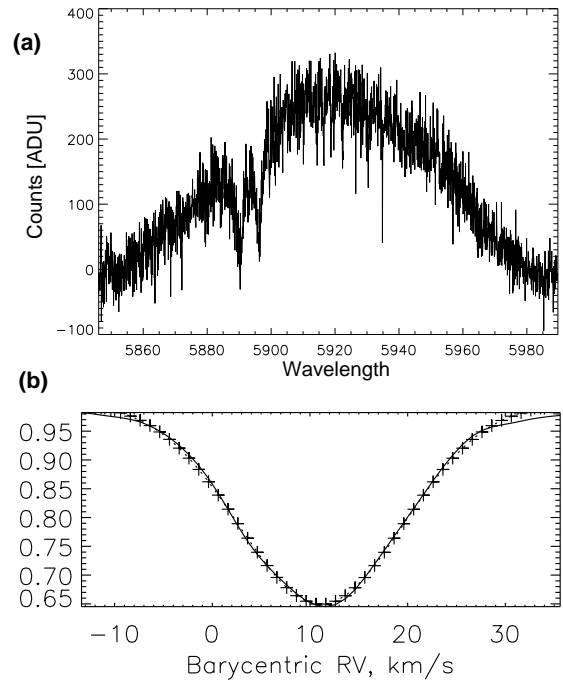}
\caption{Example  of FECH  data.   The top  plot (a) shows the  extracted
  echelle order  in the region of sodium D lines.
  The spectrum of the $V=10.5^m$  star HIP~9902B was taken on JD 2455449 with
  a 10-min.  exposure.  The lower plot (b) shows the  CCF of this spectrum
  with  the  solar mask (line)  and  its  Gaussian  fit (crosses; RV
  11.12~km~s$^{-1}$, amplitude 0.351, dispersion 8.42~km~s$^{-1}$).   The  rms
  deviation from the fit is 0.007.
\label{fig:example}
}
\end{figure}

The  FECH data  were  reduced by  the  author using  a simplified  and
adapted version of the  CHIRON pipeline software \citep{CHIRON}.  I extracted echelle
orders  84 to  123  covering  the wavelength  range  from 4500\AA  ~to
6850\AA. At shorter wavelengths, there is little signal in the spectra
of red stars and quartz lamp, while longer wavelengths are affected by
telluric absorptions and are not used in the cross-correlation.  A set
of 10 quartz spectra was co-added  and used to extract the stellar and
arc spectra in a boxcar window  of $\pm$8 pixels.  The position of the
spectra on  the CCD, defined  by the fiber,  was stable to  within 0.5
pixels.  Cosmic rays were removed during the order extraction, but
the current algorithm did not  detect all these events and some spectra
do contain a few residual spikes.

While  the  extracted CHIRON  spectra  are  corrected for  small-scale
variations  of   the  detector  sensitivity  using   the  quartz  lamp
calibration, this correction was not  applied to the FECH data because
the  quartz  ``flats'' for  FECH  deviate from  unity  by  only a  few
percent, less than the typical noise.  The FECH wavelength calibration
typically  used 1400  thorium-argon lines with a scatter  of 0.1
pixels  or 0.36\,km~s$^{-1}$ between  the measured  line positions  and their
6-th order polynomial approximation.  Each science spectrum was paired
with the  wavelength calibration closest  in time (typically  within 3
minutes).

\subsection{Radial velocities by cross-correlation}

The spectra  were cross-correlated with the digital  binary mask based
on    the     solar    spectrum stored in the NOAO archive  \citep[see][for more
details]{LCO}.
 Figure~\ref{fig:example}  shows data  on a faint  target. The
extracted  FECH  spectrum  has  $\sim$250  counts  (the  gain  is  1.4
el/ADU). However, the cross-correlation  function (CCF) is very smooth.
Its  Gaussian  fit  leads  to  $RV =  11.114  \pm  0.067$\,km~s$^{-1}$.   The
amplitude of the Gaussian is $a  = 0.351 \pm 0.002$ and its dispersion
is $\sigma  = 8.415 \pm  0.16$\,km~s$^{-1}$. The rms  residual to the  fit is
0.7\%.   The   RVs  are   derived  without  applying   any  zero-point
correction from  RV standards, they are tied directly to the solar
spectrum.

The CCF $C(v)$ is computed over the  RV range of $ v = \pm 200$\,km~s$^{-1}$.
A Gaussian  is fitted  to the  portion of the  CCF within  $\pm 2.35
\sigma$ of  the minimum. When two  or three Gaussians  are fitted, the
data  around  all  the  minima  are used  jointly.   After  the  first
iteration,  the centers  and dispersions  are determined,  and  in the
second iteration the fitting area is adjusted accordingly. The CCF
model with $k$ Gaussians is
\begin{equation}
C(v) =    1 -   \sum_{j=1}^k a_j \exp [ -(v - v_j)^2/ 2 \sigma_j^2 ] 
\label{eq:fit}
\end{equation}
and  contains $3  k$ free  parameters  $(v_j, a_j,  \sigma_j)$. It  is
possible  to   fix  some  parameters,  but  this   option,  handy  for
de-blending overlapping CCFs, was not used in the data reported below.

Formal  errors of  the fitted  parameters  do not  reflect their  real
precision,  being   dominated  by  systematic,   rather  than  random,
deviations of the CCF from its model. I do not provide these errors in
the data  table.  The  real RV precision  estimated from  residuals to
orbital fits is  about 0.1 km~s$^{-1}$ for both  CHIRON and FECH.  For
example, 4 observations  of HIP~105585C taken with CHIRON  over a time
span of 333\,days have rms scatter of 40\,m~s$^{-1}$.  The bright star
HIP~5896 with  a very wide CCF  ($\sigma=33$\,km~s$^{-1}$) observed 11
times with FECH shows the rms RV scatter of 0.28\,km~s$^{-1}$. 

For 21 apparently single stars the  RVs were measured both here and in
the  GCS.   The  average  difference  CHIRON$-$GCS  is  0.47  or  0.36
km~s$^{-1}$  (with or  without rejecting  one  outlier, respectively).
The  rms scatter of  the difference  is 0.58~km~s$^{-1}$.   This minor
systematics   is  ignored   in  the   following,  being   most  likely
attributable  to  the  CORAVEL  instruments  used by  the  GCS.   More
accurate RVs \citep{Nid02,Desidera2006} match the CHIRON data better.

\subsection{Data tables}

\begin{deluxetable*}{l rrl  | l rrl | l rrl  }    
\tabletypesize{\scriptsize}     
\tablecaption{Summary of observations
\label{tab:report} }         
\tablewidth{0pt}                                   
\tablehead{                                                                     
\colhead{HIP} & 
\colhead{RV} &
\colhead{$N$} &
\colhead{Tag} &
\colhead{HIP} & 
\colhead{RV} &
\colhead{$N$} &
\colhead{Tag} &
\colhead{HIP} & 
\colhead{RV} &
\colhead{$N$} &
\colhead{Tag} \\
& \colhead{km s$^{-1}$} &  & &
& \colhead{km s$^{-1}$} &  & &
& \colhead{km s$^{-1}$} &  & 
}
\startdata
5896       &        7.79& 10 &  1F    & 37645B     &      $-$16.39&  1 &  1C  &  98274      &      $-$19.31&  1 &  1F  \\ 
6772A      &       49.87&  1 &  1C    & 38908B     &       16.98&  1 &  1F  &  99572      &       27.27&  1 &  1F  \\
6772B      &       50.43&  1 &  1C    & 40452B     &       21.19&  1 &  1F  &  99651A     &       24.27&  2 &  1F  \\
7601       &      $-$10.61& 12 &  3CO & 41353A     &        7.31&  2 &  1F  &  100895     &      $-$14.97&  1 &  1F  \\
8486B      &       $-$6.65&  1 &  1F  & 41353B     &       28.51&  3 &  2F* &  100896     &       $-$5.81& 11 &  2F  \\
9497       &       17.81&  1 &  1F    & 44579B     &        0.60&  1 &  1F  &  101443     &       23.03&  1 &  1F  \\
9642       &       44.90& 10 &  2CO   & 45734B     &       $-$7.78&  3 &  2F  &  101551B    &       21.38&  2 &  1F  \\
9902A      &       11.48&  2 &  1F    & 45734A     &        0.14&  2 &  2F  &  101551A    &       21.41&  2 &  1C  \\
9902B      &       11.03&  2 &  1F    & 45838A     &       54.95&  1 &  1F  &  102418C    &       11.34&  1 &  1C  \\
9911B      &      $-$38.69&  2 &  1F  & 46236B     &       30.95&  2 &  1F  &  102655B    &       $-$2.34&  1 &  1C  \\
9911A      &      $-$39.13&  1 &  1F  & 47862B     &      $-$14.46&  1 &  1F  &  102655A    &       $-$2.64&  1 &  1C  \\
10579B     &       44.28&  1 &  1F    & 49030B     &       15.83&  1 &  1F  &  102945     &      $-$22.12&  1 &  1F  \\
10579A     &       38.94&  1 &  1F    & 59272B     &        2.74&  1 &  1F  &  103578     &      $-$31.01&  1 &  1F  \\
10754      &       44.82&  2 &  1C    & 59690B     &       25.10&  2 &  1F  &  104687B    &      $-$21.17&  1 &  1F  \\
11024A     &       45.19&  1 &  1F    & 60353B     &        4.35&  1 &  1F  &  104687A    &      $-$20.47&  1 &  1F  \\
11024B     &       45.19&  2 &  1F    & 60749D     &       $-$1.77&  1 &  1F  &  105569     &       $-$1.97&  1 &  2F  \\
11324A     &       34.35&  1 &  1F    & 61595B     &      $-$15.33&  1 &  1F  &  105569C    &        3.79&  1 &  1F  \\
11324B     &       32.90&  1 &  1F    & 64478A     &      $-$40.23& 10 &  3FO &  105585C    &        3.63&  4 &  1C  \\
11417      &       62.92&  1 &  2F    & 64478B     &       20.03& 22 &  2F*O&  105585A    &       $-$1.80&  4 &  2C \\
11537      &       30.23&  4 &  2F*   & 64498B     &      $-$15.73&  1 &  1F  &  105879D    &       35.90&  2 &  1F  \\
11783B     &      $-$28.26&  1 &  1C  & 64498A     &      $-$12.04&  1 &  1F  &  105879A    &       40.17&  3 &  2F*O \\
11783A     &      $-$24.58&  1 &  1C  & 66121B     &      $-$28.41&  1 &  1F  &  105947     &       22.13&  1 &  2F*  \\
11909      &        6.14&  1 &  1F    & 66676A     &        1.40&  2 &  1F  &  106438A    &      $-$29.24&  1 &  1F  \\
12326A     &       17.00&  1 &  1F*   & 66676B     &        2.37&  1 &  1F  &  106438B    &      $-$28.99&  1 &  1C  \\
12326B     &       19.21&  1 &  1F    & 67246B     &      $-$30.37&  1 &  1F  &  106632A    &       16.73&  1 &  1C  \\
12361      &       19.36&  1 &  1F    &  67408B     &        2.89&  1 &  1F  &  106632B    &       16.31&  1 &  1C  \\
12764A     &       14.75&  2 &  1F    & 69220B     &       49.95&  1 &  1F  &  107299     &       37.31&  1 &  1F  \\
12780B     &       $-$3.44&  9 &  1F*O& 71682C     &       13.22&  2 &  1F  &  107300     &       37.31&  1 &  1F  \\
12780A     &       $-$6.64&  7 &  2FO & 72235A     &        9.13&  1 &  1F  &  109035C    &      $-$14.62&  1 &  1F  \\
12884      &       43.78&  1 &  1F    & 72235B     &        9.08&  1 &  1F  &  109035B    &        4.74&  1 &  1F  \\
13498      &       13.88&  8 &  3CO   & 74975B     &       55.08&  1 &  1F  &  109035A    &       $-$8.67&  1 &  1F  \\
13725      &       10.28&  1 &  1F    & 76435C     &        7.34&  2 &  1F* &  109951     &      $-$24.04&  1 &  2F*  \\
14194A     &       35.32&  1 &  1C    & 76435A     &        5.47&  3 &  1F  &  110091     &       16.05&  1 &  1F  \\
14194B     &       31.86&  3 &  2C*   & 76888A     &        8.06&  1 &  1F  &  110447A    &        2.47&  1 &  1C  \\
14307B     &       19.89&  3 &  2C*   & 76888B     &        7.90&  1 &  1F  &  110447B    &        1.37&  1 &  1C  \\
14519B     &       17.70&  2 &  1C    & 79730B     &      $-$40.08&  1 &  1C  &  110712A    &       14.57&  1 &  1F  \\
14519A     &       15.87&  2 &  1C*   & 83701B     &       $-$1.09&  1 &  1C  &  110712B    &       15.28&  1 &  1F  \\
16860B     &     $-$24.93&  1 &  1C   & 85342B     &      $-$15.85&  1 &  1C  &  111903     &       41.68&  1 &  1F* \\
22611B     &       46.03&  1 &  1C    & 87813B     &       $-$5.94&  3 &  1C  &  112201B    &      $-$12.87&  1 &  1C  \\
22611A     &       45.72&  1 &  1C    & 89805B     &        1.61&  1 &  1F  &  112325     &      $-$28.07&  1 &  1F  \\
22611C     &       46.39&  1 &  1C    & 89805A     &        0.87&  1 &  1F  &  113386A    &        0.46&  1 &  1F  \\
22683B     &        4.71&  1 &  1C    & 89808      &        8.14&  1 &  1F  &  113386B    &        0.85&  2 &  1F  \\
22693B     &      $-$0.65&  1 &  1C   & 91837C     &       $-$3.65&  1 &  1F  &  113579     &        7.03&  1 &  1F  \\
23926B     &       45.09&  1 &  1F   & 91837A     &       $-$3.58&  1 &  1F  &  113597     &       20.55&  5 &  3F* \\
24711B     &     $-$10.79&  1 &  1C   & 92140      &        4.60&  9 &  1F  &  114167B    &        4.81&  1 &  1F  \\
25082B     &       23.23&  1 &  1F  & 94310      &      $-$15.12&  1 &  1F  &  114167A    &        4.25&  1 &  1F  \\
27922B     &       44.03&  1 &  1F  & 94389      &       23.49&  1 &  1F  &  115087     &       15.21& 10 &  1FO\\
28790B     &       34.92& 10 &  2C*O& 95106A     &        9.35&  2 &  1F  &  116063B    &        4.30&  1 &  1F  \\
28790A     &       17.20&  8 &  1CO & 95106B     &        9.29&  5 &  1F* &  116063A    &        4.03&  1 &  1F  \\
31711C     &       31.58&  1 &  1C  & 95116B     &      $-$41.12&  3 &  1C  &  117081A    &       21.79&  1 &  1F  \\
35261C     &       23.83&  3 &  2C  & 95116A     &      $-$41.93&  1 &  1C  &  117081B    &       21.24&  1 &  1F  \\
36165B     &       64.86&  1 &  1C  & 95847      &      $-$23.95&  1 &  1F  &  117391B    &       17.74&  1 &  1C  \\
36165A     &       66.23&  1 &  1C  & 97508C     &        8.04&  2 &  1F  &  117666     &        9.82&  1 &  2F* \\
36640B     &       56.25&  1 &  1C  & 97508A     &        8.32&  1 &  1F  & &   &  & \\
37645A     &     $-$19.52&  1 &  1C  & 97548      &       $-$8.46&  1 &  1F  &  &   &  & 
\enddata 
\end{deluxetable*}

Table~\ref{tab:report}  lists all  163 observed  targets.   Column (1)
identifies  the target  by  the  {\it Hipparcos}  number  of the  main
component and,  where necessary, the  component letter.  As  most (but
not all)  targets belong to  the FG-67 sample,  additional information
such  as equatorial  coordinates,  proper motions,  and magnitudes  in
various  pass-bands  can  be  retrieved  from the  FG67a  paper.   The
following columns list the average RV, the number of observations, and
the tag that combines the  maximum number of fitted Gaussians (e.g.  2
for a double-lined binary) with  the instrument code, C for CHIRON and
F for  FECH.  Asterisk  after the tag  signals a  spectroscopic system
discovered in this survey, the letter O means that spectroscopic orbit
has been computed.
 
Table~\ref{tab:RV}, published in full electronically, gives individual
observations.   Its  column  (1)  contains  the  target  name  (as  in
Table~1).  Columns (2) to (5) contain the heliocentric Julian date and
the parameters $(v,a,  \sigma)$ from the Gaussian fits (equation \ref{eq:fit}).   If more than
one component was fitted,  the parameters of additional components are
listed separately, with the same  date. The last column (6) contains a
tag  that combines  the number  $j$ of  the fitted  Gaussian  with the
instrument code.   The Table contains 343  individual observations, 233
made with FECH and 110 with CHIRON.

\begin{deluxetable}{l c ccc l}    
\tabletypesize{\scriptsize}     
\tablecaption{Individual radial velocities (fragment)
\label{tab:RV}          }
\tablewidth{0pt}                                   
\tablehead{                                                                     
\colhead{HIP,} & 
\colhead{HJD} &
\colhead{RV} &
\colhead{$a$} &
\colhead{$\sigma$} &
\colhead{Tag} \\ 
\colhead{Comp.} &
\colhead{+24\,00000} & 
\colhead{km s$^{-1}$} &
&
\colhead{km s$^{-1}$} &
}
\startdata
 5896     &     55490.6235 &  7.784 &  0.040 & 33.504 & 1F \\
 5896     &     55510.5567 &  7.880 &  0.040 & 33.680 & 1F \\
 6772A    &     56863.8045 & 49.866 &  0.201 &  5.638 & 1C \\
 6772B    &     56863.7976 & 50.427 &  0.280 &  6.495 & 1C \\
 8486B    &     55436.9363 &$-$6.649&  0.299 &  4.973 & 1F \\
 9497     &     55436.9452 & 17.811 &  0.223 &  5.318 & 1F \\
 9902A    &     55449.8915 & 11.836 &  0.072 & 18.887 & 1F
\enddata 
\end{deluxetable}

\begin{deluxetable*}{l l }    
\tabletypesize{\scriptsize}     
\tablecaption{Comments on individual objects
\label{tab:disc} }         
\tablewidth{0pt}                                   
\tablehead{                                                                     
\colhead{HIP} & 
\colhead{Comment} 
}
\startdata
6896 &  F6IV star belongs  to the nearby (20\,pc) visual quadruple system $\kappa$~Tuc. 
         The RV is  1.1 km~s$^{-1}$ and variable (GCS),  \\ 
     & 7.8  km~s$^{-1}$ and constant here.  {\it Hipparcos} acceleration. Orbital period of several years? \\
7601 & SB3, two orbits and speckle resolution, periods 19.4 and 614 days. \\
9497 & Close visual triple system without prior RV data.  The CCF has only one narrow dip. \\
9642  & SB2 discovered in GCS, orbit $P= 4.78$\,d. \\
10579 & Algol-type eclipsing binary DS~Cet, but the CCFs of A and B  are narrow. \\
10754 & The component A = HIP~10621 at 836\arcsec is optical, different RV. \\
11417 & SB2 discovered by D.~Latham (2012, private communication);  here  double-lined with equal components. \\
11537 & * New SB2, {\it Hippparcos} acceleration. Possibly resolvable
by speckle interferometry. \\
11783 & A was resolved  at SOAR at 0\farcs2, which explains its RV difference with B. \\
12326 & * A has asymmetric CCF, RV differs from B and C. Possibly
resolvable by  speckle. \\
12780 & * B is new SB1, orbit $P=27.8$\,d. A is SB2 and visual binary.  \\
12884 & Eclipsing  binary CN~Hyi  in a  78-year visual binary. The RV here refers to the visual secondary with narrow CCF. \\  
&   Two wide, low-contrast  dips of the  eclipsing pair are also present in the CCF, but they were not measured. \\
13498 &  SB3 discovered by \citet{LCO}, orbit with $P=13.75$\,d, correction of the visual orbit.  \\
14194 & * B is new SB2 with a period on the order of a year. \\
14307 & * B has asymmetric CCF. It was resolved at SOAR at 0\farcs19, estimated period 25\,yr. \\
14519 & * RV difference of 1.8~km~s$^{-1}$ between A and B  indicates a subsystem in one of the components. \\
22683 & The B-component HIP~22668 at 796\arcsec has different RV, it is either optical  or contains a subsystem. \\
26611 & Three  components at  100\arcsec and 52\arcsec in  non-hierarchical configuration have matching RVs. \\
28790 & * Orbits for A ($P=221$\,d) and B (new SB2, $P=13.23$\,d).  \\
35261 & C is an SB2 discovered by \citet{LCO}. \\
37645 &  A is a known SB resolved  at SOAR at 0\farcs20, hence the RV difference with B. \\
41353 & * B=HD~71842 is a new SB2, at 297\arcsec from A, optical. For
this reason orbit determination is not attempted.  \\
45734 & Young quadruple:  A is a  close    pair,   while    B    has   double    lines \citep{Desidera2006}. \\
      & The CCFs of  B are of strange rectangular shape, possibly triple-lined, while the CCF of A is asymmetric. \\
64478 & * B is a new SB2, orbit $P=0.24$\,d. Updated orbit of A, $P=4.3$\,d \\
66676 &  B is a triple system resolved at SOAR, but its CCF is narrow, while the RVs of A and B match. \\
72235 &  B was resolved at SOAR at 0\farcs4, but its CCF is narrow, while the RV matches that of A. \\
76435 & * RV(C) differs from RV(A). C was resolved at SOAR at 56\,mas, estimated period 4\,yr. \\
85342 &  B = HIP~85326 is a spectroscopic and acceleration binary resolved at SOAR. \\
87813 &  B is a spectroscopic binary discovered by \citet{TS02}. \\
92140 & SB1 with slow RV variation  (FECH, also N.~Gorynya, private commuication, 2013).  \\
95106 & * B = HIP~95110 is a new SB1 with period of a few months. A was resolved at SOAR at 0\farcs27.  \\
95116 & HIP~96979 has matching RV,  similar proper motions and parallaxes. Too distant to be a bound binary, however.\\
95708 &  AB is a 65-year visual binary, its CCF is asymmetric. The RV(C) is similar, physical triple. \\
98274 &  Close visual binary, its CCF is asymmetric. This is the first RV measurement. \\
100896 & The RVs match the 160-day SB2 orbit by  \citet{Halb2012}, B  = HIP~100895 is physical.\\
102945 & Fast axial rotation ($\sigma=20.2$\,km~s$^{-1}$); it is   a 0\farcs8 visual binary with evolved component. \\ 
105585 & AB is a visual binary I~337, seen as SB2 here. All data can be fitted by a 207-year orbit with $e=0.72$. \\
105879 & * A is SB2, preliminary orbit with $P=9$\,yr, also {\it Hipparcos} acceleration. \\
105947 & * SB2 and 20.7-year visual binary with a well-defined orbit. $\Delta V =17$~km~s$^{-1}$  is 2.2 times larger \\
             & than predicted, caused by a subsystem? \\
109951 & * Double CCF, suspected subsystem in the secondary component. \\
             & $\Delta V=12.5$~km~s$^{-1}$ is 3 times larger than     predicted by the visual orbit. \\
110091 & Close visual binary CHR~107 with estimated period of $\sim$6\,years. The CCF is slightly asymmetric. \\
110792 & B=HIP~110719 at  20\farcs6 was resolved at SOAR at  0\farcs19, explaining the RV difference. \\
111903 & * RV(A) differs by 3.4 km~s$^{-1}$ from the RV reported in GCS, hence new SB?  \\
113597 & * New SB3.  It is a 1\farcs8 visual binary, while HIP~113579 at 581\arcsec also belongs to the system.  \\
115087 &  SB1 discovered by GCS, orbit $P=7.88$\,d. \\
117666 & * The 30-year visual orbit of A~2700 predicts $\Delta V$ of
2.7~km~s$^{-1}$, here  $\Delta V = 16.3$~km~s$^{-1}$, hence a subsystem. 
\enddata 
\end{deluxetable*}

Short comments on new spectroscopic systems or other objects for which
the new  data are interesting  in some or  other way are  assembled in
Table~\ref{tab:disc}.   Asterisks   mark  17  systems   or  subsystems
discovered  in  this  survey.   Orbital  periods are  given  here  for
reference only, pending publication  of the full analysis.  Some stars
with   asymmetric  CCFs,  presumably   binaries  with   components  of
comparable luminosity and small  RV difference, were resolved recently
by speckle interferometry at the SOAR telescope \citep{SAM15}.

\section{Detection of subsystems from  RV difference between binary components}
\label{sec:simul}

Subsystems  with  large  mass  ratios  are  readily  detected  by  the
appearance of double or blended  lines in a single spectrum.  When the
visual orbit is known, the RV difference for the moment of observation
can be computed using estimated  masses. If the observed RV difference
is discrepant  even after re-adjusting the visual  orbit, the presence
of a subsystem in one of the components is strongly suspected. This is
the case of HIP 105947, 109951, and 117666.

The alternative idea of detecting subsystems in wide binaries from the
RV difference between  the components (as in HIP  12780, 14519, 76435)
is explored  here by means of  simulation.  We need  to decide whether
the  wide binary contains  a subsystem  (hypothesis H1)  or is  just a
binary  (hypothesis H0),  based  on  a single  measurement  of the  RV
difference  between its  components $\Delta  V  = |V_1  - V_2|$.   The
quantity $\Delta  V$ is measured  with some error. Moreover,  the wide
binary is itself  in orbital motion, so the RVs  of its components are
not exactly equal even without subsystems.

First,  I   establish  the  distribution   of  the  orbital   RVs  of
spectroscopic  binaries. The  RV semi-amplitude  $K_1$ depends  on the
orbital inclination $i$ and eccentricity $e$ as
\begin{eqnarray}
K_1 & = &  A_1 \sin i (1 - e^2)^{-1/2}, \nonumber \\
A_1 & = &  29.8 \; P^{-1/3} M_2 (M_1 + M_2)^{-2/3} .
\label{eq:K1}
\end{eqnarray}
Here $A_1$ (the RV semi-amplitude in a circular orbit at $i=90^\circ$,
in km\,s$^{-1}$) is  related to the orbital period  $P$ (in years) and
the  component masses  $M_1$  and $M_2$  (in  solar-mass units).   The
coefficient 29.8\,km~s$^{-1}$ is the average orbital velocity of the Earth.

A large number ($N=10\,000$)  of spectroscopic binaries was simulated.
Their RV $V_1$ was calculated  at random orbital phase.  The influence
of masses  and periods is  eliminated by studying the  distribution of
dimensionless  quantity $x  =|V_1|/A_1$.  Only  the random  effects of
inclination, eccentricity, and phase remain.  I assume isotropic orbit
orientation which corresponds to the uniform distribution of $\cos i$.
The  distribution  of  $x$  depends  on the  distribution  of  orbital
eccentricity. Four cases are  considered: (1) $e=0$ (circular orbits),
(2)  sine distribution  $f(e)= \pi/2  \;  \sin (\pi  e)$, (3)  uniform
distribution   $f(e)=1$, and  (4) linear distribution  $f(e)=2e$,
often called ``thermal''. In
the additional  case 2a, the eccentricity  of case 2  is multiplied by
0.7 to get the average $e=0.35$ typical of spectroscopic binaries.

\begin{figure}
\plotone{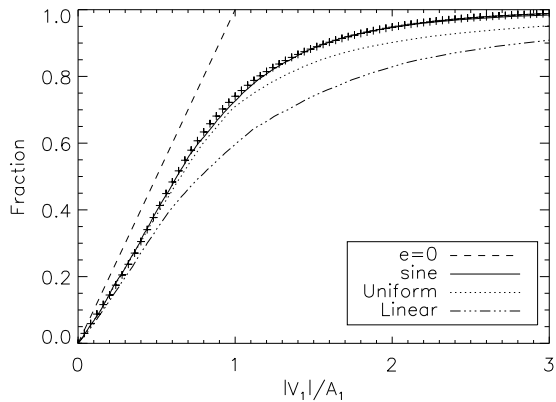}
\caption{Cumulative distributions of normalized orbital velocities 
$x =|V_1|/A_1$ for different
  eccentricity distributions of spectroscopic binaries. The plus signs
  correspond to the analytical approximation. 
\label{fig:simul} 
}
\end{figure}

Figure~\ref{fig:simul} shows the results.  In the case (1), the strict
inequality  $K_1 \le  A_1$  holds.  Combination  of  random phase  and
random inclination results  in the uniform distribution of  $x$ in the
interval $[0,1]$,  so its cumulative  distribution is linear,  $F(x) =
x$.   Larger RV variations  are produced  in eccentric  orbits.  Their
probability  is  small, increasing  with  the  increasing fraction  of
eccentric  orbits.   On   the  other  hand,  the  median   of  the  RV
distribution depends on the eccentricity much less.

A spectroscopic binary can be detected by a large RV variation. The
probability of such event $p_{\rm det}(x) = 1 - F (x)$ is defined by the
cumulative distribution $F(x)$. I found a good analytical approximation for
the case 2, 
\begin{equation}
p_{\rm det}(x) = 1 - F (x)  \approx  10^{ -x[0.65 - 0.32/(1 + 4 x^3)] } 
\label{eq:Fx}
\end{equation}
(crosses  in   Figure~\ref{fig:simul}).   If  the  eccentricity  is
distributed  uniformly (case  3), the  formula (\ref{eq:Fx})  is still
good at small $x$.

The observed RV difference $\Delta V$ includes a contribution from the
orbital  motion in  wide binary.   A  subsystem is  detectable from  a
single observation  when $\Delta V$ significantly  exceeds the orbital
velocity of  the outer binary.  The latter  is estimated statistically
using simulations.  The relevant normalization factor is now
\begin{equation}
A_L =   29.8 \; P_L^{-1/3} M_L^{1/3} 
\label{eq:A0}
\end{equation}
instead of $A_1$.  I denote the  ``long'' period of the wide binary by
$P_L$, its mass  sum by $M_L$, and the  corresponding amplitude of the
RV difference  by $A_L$.  For the  long period of  $10^3$\,years and a
mass  sum of 1~${\cal  M}_\odot$, the  amplitude is  0.1 of  the Earth
speed or 3\,km~s$^{-1}$.  Considering  that the errors of RVs measured
with CHIRON are on the order of 0.1\,km~s$^{-1}$, it is clear that the
RV difference is dominated by the motion in wide binaries, while the
measurement errors are negligible in comparison.

\begin{figure}[ht]
\plotone{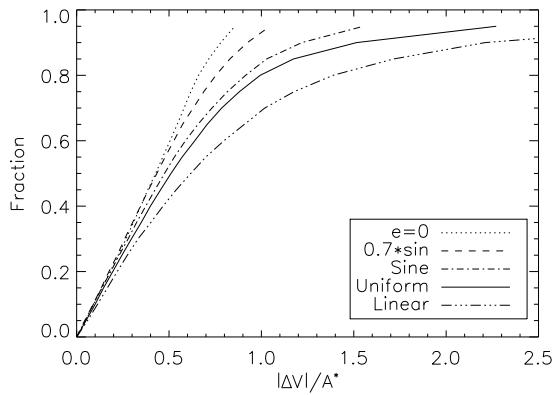}
\caption{Cumulative  distributions of normalized orbital
  RV difference $F(|\Delta V|/A^*)$  for wide  binaries
  with various eccentricity distributions. 
\label{fig:sim5} 
}
\end{figure}

In  the case  of wide  binaries,  we do  not know  their true  orbital
periods $P_L$ and estimate  them by assuming that projected separation
equals   semi-major  axis  $a_L$.    Considering  that   $P_L  \propto
a_L^{3/2}$,   the    estimated   amplitude   becomes    $A^*   =   A_L
(a_L/\rho)^{0.5}$.  This  has been  accounted for in  the simulations.
Normalized RV amplitudes using projected separations ($A^*$ instead of
$A_L$) are  slightly smaller  because, statistically, $A^*$  is larger
than $A_L$.   Cumulative distributions for different  cases are plotted
in Figure~\ref{fig:sim5}.

The  probability  that  the  measured  RV  difference  $\Delta  V$  is
compatible with the hypothesis H0 (no subsystems) is estimated through
cumulative distribution $ F( |\Delta V|/A^*)$ as
\begin{equation}
p({\rm H0}) = 1 - F( |\Delta V|/A^*). 
\label{eq:PH0}
\end{equation}
Let  $\epsilon$  be a  small  number  characterizing  the false  alarm
probability  (detection of  a  non-existent subsystem).  The
hypothesis H0  is rejected if $ p({\rm H0})  < \epsilon$, or $ F( |\Delta V|/A^*) >
1 -  \epsilon$.  Adopting $\epsilon  = 0.05$, the detection  limit for
subsystems  translates  to  $x>k_L(\epsilon)$,  where  $k_L=0.85$  for
circular  outer   orbits,  $k_L=  1.55$  for   the  sine  eccentricity
distribution,   and  $k_L=2.40$  for   the  uniform   distribution  of
eccentricity in the outer orbits.

Now the tools are in place  to estimate the probability of detecting a
subsystem  in a  wide  binary  from a  single  measurement of  $\Delta
V$. For each  combination of the mass ratio $q_S$  and period $P_S$ in
the subsystem,  the amplitude $A_1$ is computed  by (\ref{eq:K1}). The
detection threshold is $V_1 > k_L A^*_L$ or $ x > k_L A^*_L /A_1$. The
probability  of such  event is  estimated by  the  approximate formula
(\ref{eq:Fx}).   A  single measurement  of  $\Delta  V$ provides  thus
statistical  constraints  on  the   presence  of  subsystems  in  both
components of  a wide binary.  This  reasoning does not  apply to wide
binaries with  known subsystems,  e.g.  containing close  inner visual
binaries,  where $\Delta  V$ is  likely  produced by  motion in  those
subsystems, rather than in the wide binary itself.

\begin{figure}[ht]
\plotone{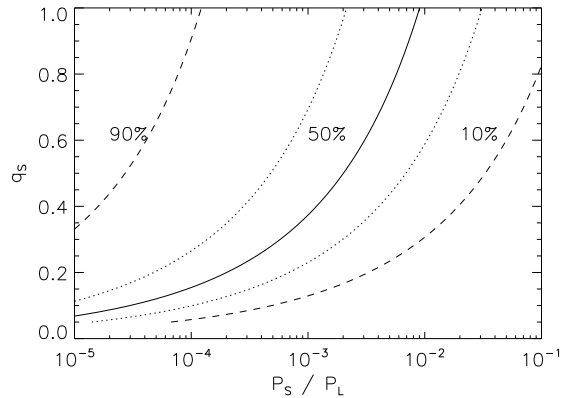}
\caption{Probability  of detecting  a subsystem  by the  RV difference
  method as a function of the period ratio $P_S/P_L$ and mass ratio in
  the  subsystem  $q_s$.   The  curves,  left  to  right,  depict  the
  detection  probability of 90\%,  70\%, 50\%  (full line),  30\%, and
  10\%.  
\label{fig:bindet} 
}
\end{figure}

Period ratio  vs.  mass ratio  in detectable subsystems is  plotted in
Figure~\ref{fig:bindet} with the following assumptions: $M_L = 2 M_1$,
sine eccentricity  distribution in inner subsystems, and  $k_L = 2.4$.
Note the  ``fuzzy'' character of this detection  technique: the chance
to miss  a subsystem is substantial  for all parameters.   This is the
penalty for using just a single RV measurement. The detection curves
are described by
\begin{equation}
\frac{q_S}{(1 + q_S)^{2/3}} > \frac {k_L}{x(p_{\rm det})} \; (2P_s/P_L)^{1/3} ,
\label{eq:crit2}
\end{equation}
where  $x(p_{\rm  det})$ is  determined  by inverting  (\ref{eq:Fx}),
namely 0.15, 0.39, 0.63, 0.95, and 1.60 for probabilities of 0.1, 0.3,
0.5, 0.7, and 0.9 respectively.

Subsystems  with  $q_S  >  0.75$  can  be  detected  by  double  lines
independently  of the outer  period. Several  detections in  this work
belong to this  category.  However, doubling of the  CCF is observable
when  the  RV  difference  in   the  inner  subsystem  exceeds  a  few
km~s$^{-1}$, in which case the single-measurement strategy also works.
For simplicity, I do not account for this additional detection channel
in the statistical analysis of the next Section.

\section{Statistics of subsystems in wide binaries}
\label{sec:widebin}

The list  of wide binaries in  the FG-67 sample was  created using the
following criteria:  separation larger  than $8''$, $V$  magnitudes of
the secondary from 8 to 11, declination south of $+20^\circ$.  The
original list of 112 binaries was cleaned from optical systems and one
white dwarf companion, leaving 98 wide binaries.

Table~\ref{tab:widelist}  concentrates  relevant  information  on  the
selected binaries.   Its column (1)  is the {\it Hipparcos}  number of
the primary component and  designations of the wide-binary components.
Columns (2)  to (5)  give the parallax  \citep[mostly from][]{HIP2}, the
visual   magnitudes   of   both   components,  and   the   separation,
respectively.  Column (6) contains the characteristic amplitude of the
orbital motion  in the wide  binary $A^*$ computed from  the projected
separation  and masses  of the  components  given in  FG67a.  Then  in
columns (8) and  (9) the average RVs  of the primary and
secondary components are given, while Column (9) gives reference codes
for these  RVs, explained in  the notes to  the table.  When  only one
code is given, it covers both components.  In two cases I found only
low-accuracy RVs in  SIMBAD.  If there are several  RV sources, I gave
preference  to this  work and/or  selected  the same  source for  both
components  where   available.   The  average   RVs  of  spectroscopic
subsystems are replaced by 'S2' or S1'.  The presence of subsystems in
the primary  and secondary components  of each binary is  indicated in
 Columns  (10) and  (11), respectively, using  the codes  of FG67a.
Briefly, 'C' means subsystems wider than $\sim 3''$, 'V' and 'v' stand
for  close  resolved  subsystems,  's2'  and  'S2'  mean  double-lined
spectroscopic binaries,  's' and 'S1' mean  single-lined binaries, 'a'
stands   for   acceleration   binaries.    A   plus   sign   indicates
sub-subsystems  (spectroscopic   pair  in  the   close  visual  binary
HIP~64478A).   The last  Column (10)  gives {\it  Hipparcos}  or other
identifications of secondary components, where available.

Not  all subsystems  are  absolutely certain,  e.g. the  spectroscopic
subsystems in both components of HIP~36165 announced in the GCS.  I do
not  include the potential  subsystem in  HIP~14519 discovered  by the
$\Delta V$  method.  The apparently  non-hierarchical system HIP~22611
is removed from the following analysis, leaving 96 wide pairs.

Subsystems are found  in 43 pairs, 25 of which  have subsystems in the
primary  component,  28  in the  secondary,  and  10  in both  ($43  =
25+28-10$). Obviously, the frequency  of subsystems in the primary and
secondary components  is statistically the same.  This  sample gives a
weak  evidence of  the  correlated occurrence  of  subsystems in  both
components,  noted  previously  in   \citep{RAO}  and  FG67b.   If  the
occurrence of subsystems  were statistically independent, the expected
number of  wide binaries with  subsystems in both components  would be
$96  \times (25/96)  \times (28/96)  =  7.3$, slightly  less than  the
actual number 10.

\begin{figure}[ht]
\plotone{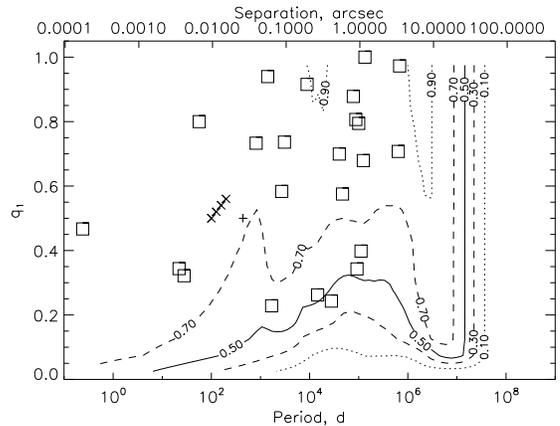}
\caption{Secondary subsystems in  the $(P,q)$ space (squares). Crosses
  denote  unknown periods  and mass  ratios, placed  arbitrarily.  The
  contours show  average detection  probability. The upper  axis gives
  approximate angular separation at a distance of 50\,pc.
\label{fig:sec}
}
\end{figure}

Figure~\ref{fig:sec} compares periods and mass ratios of subsystems in
the  secondary  components  with  their  detection  probability.   The
probability  is calculated  by the  prescription of  FG67a.  The
additional   factor  resulting   from  $\Delta   V$   measurements  is
incorporated  for  the  53  binaries  without  subsystems.   The  four
subsystens  with  unknown periods  and  mass  ratios  are depicted  by
crosses located  arbitrarily.  The addition  of spectroscopic coverage
has dramatically improved the detection at $P<10^4$ days in comparison
with  imaging-only surveys  such as  \citep{SAM}.  Such  plot  for the
primary  subsystems   is  similar,  with   slightly  deeper  detection
contours.

\begin{figure}[ht]
\plotone{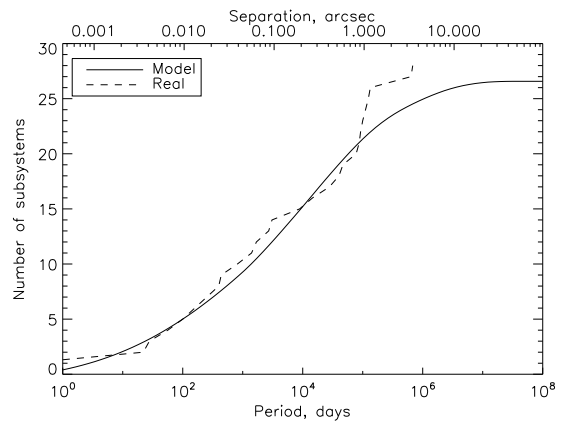}
\caption{Cumulative  distribution of  periods in  secondary subsystems
  (dashed line) and its model for $\beta=1$ (full line).  
\label{fig:cum}
}
\end{figure}

It is instructive  to compare the frequency and  periods of subsystems
with the statistical model developed in FG67b. The basic assumption is
that subsystems  are chosen randomly from  the log-normal distribution
of periods, keeping only  dynamically stable multiples.  I adopted the
following  parameters of  the underlying  period distribution:  $x_0 =
4.54$,  $\sigma= 2.40$,  the  mass-ratio exponent  $\beta=1$, and  the
total binary frequency $\epsilon =  0.5$ (see FG67b for description of
those parameters).  Two-dimensional  distribution in the $(P,q)$ space
was  multiplied  by  the   detection  probability  and  the  dynamical
truncation,  providing   the  expected  period   distribution  in  the
subsystems  and their  total number.   It  is compared  to the  actual
cumulative distribution of  secondary periods in Figure~\ref{fig:cum}.
The  four yet  unknown periods  were given  arbitrary values  close to
100\,days.   The   model  predicts  26.6   sevcondary  subsystems  for
$\beta=1$ and 23.1 for $\beta=0$, while the actual number is 28.  This
calculation  was repeated  for  the primary  subsystems, giving  their
predicted  numbers  of 32.4  and  29.1  for  $\beta=1$ and  $\beta=0$,
respectively;  the actual  number  is 25.   The  numbers match  within
statistical errors, as do the period distributions (with  the caveat related
to unknown  periods).  Therefore, the data collected  here support the
statistical description of multiplicity proposed in FG67b.


\section{Eccentricities of wide binaries}
\label{sec:ecc}

\begin{figure}[ht]
\plotone{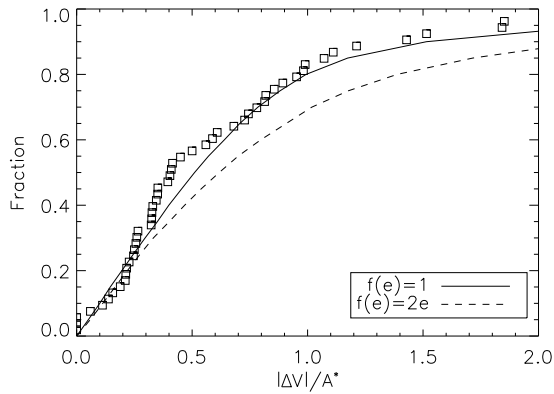}
\caption{Cumulative distribution of the normalized RV difference in 53
  wide  binaries  without  subsystems  (squares) is  compared  to  
  simulations    for    uniformly   distributed    outer
  eccentricities  (full line)  and for  the linear  outer eccentricity
  distribution (dashed line).
\label{fig:dv}
}
\end{figure}

In  this  Section,  I  study  the  distribution  of  the  observed  RV
difference  in  53  wide   binaries  without  known  subsystems.   The
cumulative distribution of the normalized quantity $|\Delta V|/A^*$ is
plotted   in  Figure~\ref{fig:dv}  and   compared  to   the  simulated
distributions  presented in  Figure~\ref{fig:sim5}.  A  good agreement
with the case-3 simulation  (uniformly distributed eccentricity) and a
poor match with the ``thermal'' eccentricity distribution are obvious.
The observed distribution is widened by the RV errors (both random and
systematic)  and   by  yet   undetected  subsystems.   So,   the  real
distribution should be even narrower than the observed one. This means
that the eccentricities of wide  binaries are likely even smaller than
in the case 3.

\citet{Reipurth2012}  suggested  that  very  wide  binaries  can  form
through  ejection of  one  star from  compact  triple systems.   Outer
orbits  of such ejected  stars mostly  have large  eccentricities with
typical values  of $\sim$0.7 (see their Figure~2).   The wide binaries
studied here  have separations on the  order of $10^3$\,AU,  as in the
above  paper.   However,  eccentricities  of  those  binaries  without
subsystems appear to  be moderate, as follows from  the RV differences
between their components. Moreover, the frequency of subsystems in the
primary and secondary components  is similar, which speaks against the
ejection    scenario.    Interestingly,    \citet{R10}    found   that
eccentricities  of  binaries  in  the  25-pc  volume  are  distributed
approximately uniformly independently  of their period.  However, they
based this  conclusion on known spectroscopic and  visual orbits which
do not include binaries as wide as studied here.

\section{Summary}
\label{sec:sum}

Results of the  RV survey of nearby stars are  reported.  I focused on
wide  visual  binaries and  aimed  at  detecting  subsystems in  their
components  by a  presence  of double  lines  or by  a substantial  RV
difference between the components.  This latter ``snapshot'' technique
is studied by  simulation, showing that the major  factor limiting the
subsystem detection is  the orbital motion in the  wide binary itself.
The   survey   readily   detects   subsystems  with   short   periods.
Non-detections   are  interpreted   statistically,   constraining  the
parameters of subsystems and their overall frequency.

As most of the primary components  of  wide binaries in this sample
already have  RV coverage in the literature, I observed fainter and
less-well studied  secondaries. Five  new subsystems in  wide binaries
were discovered in  this project.  The frequency of  subsystems in the
primary  and secondary components  turns out  to be  statistically the
same.  The  data  match  the  crude model  proposed  in  FG67b,  where
subsystems are  chosen randomly  from the general  period distribution
and are constrained only by dynamical stability.

An  interesting  additional  result  concerns  wide  binaries  without
subsystems.  The  distribution of  the observed RV  difference between
their components  places some constraints  on the distribution  of the
orbital eccentricities,  which are not measured directly  owing to very
long  periods. The  results  of  this work  rule  out the  ``thermal''
distribution $f(e)  = 2e$ and match instead  the uniform distribution.
This suggests that dynamical  interactions in $N$-body stellar systems
were not dominant in shaping the orbits of wide binaries.

\acknowledgements

I  thank   the  operators  of   the  1.5-m  telescope   for  executing
observations  of  this  program  and  the  SMARTS  team  at  Yale  for
scheduling and pipeline processing.  This work used the SIMBAD service
operated  by  Centre des  Donn\'ees  Stellaires (Strasbourg,  France),
bibliographic references from  the Astrophysics Data System maintained
by  SAO/NASA, and  the Washington  Double Star  Catalog  maintained at
USNO.

{\it Facilities:}  \facility{CTIO:1.5m}.





\clearpage

\LongTables


\begin{deluxetable}{l   r rr rr  rr l  l l l }                                 
\tabletypesize{\scriptsize}     
\tablecaption{Subsystems in wide binaries
\label{tab:widelist} }                               
\tablewidth{0pt}                                   
\tablehead{                                                                     
\colhead{HIP,} & 
\colhead{$\pi_{\rm HIP}$} &
\colhead{$V_1$} &
\colhead{$V_2$} &
\colhead{Sep.} &
\colhead{$A^*$} &
\colhead{RV1} &
\colhead{RV2} &
\colhead{Reference} &
\colhead{Sys1} &
\colhead{Sys2} &
\colhead{Comment}
\\
\colhead{Comp.}   & 
\colhead{(mas)} &
\colhead{(mag)} &
\colhead{(mag)} &
\colhead{($''$)} &
\colhead{(km~s$^{-1}$)} &
\colhead{(km~s$^{-1}$)} &
\colhead{(km~s$^{-1}$)} &
 &
 & 
}
\startdata 
  2292 A,B&   17.6 &    7.9 &    9.4 &  839.3 &    0.2 & 9.6  & 9.6   & GCS,L02     & \ldots & v & B=2350\\
  2888 A,B&   22.9 &    6.8 &    8.5 &  329.8 &    0.5 & S2  & S1   &  SB9,T06     & S2 & S1 & B=2848 \\
  3203 A,B&   37.7 &    7.0 &    9.9 &    8.5 &    2.6 & 12.0 & n/a   & N04,CfA    & \ldots & \ldots & \\
  3795 AB,C & 15.6 &    7.7 &   10.8 &  152.3 &    0.6 & --21.40 & n/a &   GCS       & V & v & \\
  6772 A,B&   16.6 &    8.2 &   11.1 &  389.1 &    0.3 & 49.9 & 50.4  & CHI         & \ldots & \ldots & B=6804 \\
  9902 A,B&   22.6 &    7.6 &   10.5 &   52.2 &    0.9 & 11.5 & 11.0  & CHI         & \ldots & \ldots & \\
  9911 A,B&   27.1 &    7.0 &   10.5 &   79.2 &    0.8 & --39.1& --38.7 & CHI     & \ldots & \ldots & \\
 11024 A,B&   24.4 &    8.0 &   10.4 &   10.3 &    1.9 & 45.2 & 45.2  & CHI     & \ldots & \ldots &\\
 11137 A,B&   17.0 &    8.9 &    9.4 &   34.7 &    0.9 & 26.3 & 26.6  & L02      & \ldots & \ldots & B=11134 \\
 11783 A,B&   37.5 &    4.8 &    8.8 &  345.3 &    0.5 & --24.6 & --28.3 & CHI     & s,a,v & \ldots & B=11759 \\
 12780 A,B&   24.2 &    7.0 &    8.5 &   12.5 &    2.4 &  S2 & S1     & CHI        & V,S2 & S1 & \\
 14194 A,B&   18.8 &    7.6 &   10.0 &    8.7 &    2.2 & 35.3    & s2 & CHI       & \ldots & s2 & \\
 14307 A,B&   18.7 &    7.6 &    8.6 &   38.3 &    1.1 & 20.4   &  s2  & GCS,CHI  & \ldots & v,s2 & B=14313 \\
 14519 A,B&   18.8 &    9.1 &   10.6 &   60.3 &    0.7 & 15.9   & 17.7 & CHI      & \ldots & s1? & \\
 15527 A,B&   28.1 &    7.4 &    8.5 &  253.0 &    0.4 & 39.5  &  39.6  & GCS     & \ldots & \ldots & B=15526 \\ 
 16860 A,B&   21.1 &    9.0 &   10.9 &   31.5 &    1.0 & --26.1 & --24.9  & GCS,CHI & \ldots & \ldots & B=16858 \\
 18888 A,B&   15.9 &    8.2 &    8.6 &  595.2 &    0.3 & 33.7  & 33.7  & GCS      & \ldots & v & B=18958 \\
 21923 A,B&   23.0 &    7.2 &   10.2 &  141.6 &    0.6 & 14.5:  & S2  & SIM,H12      & \ldots & S2 & B=21946 \\
 22611 A,C &  16.9 &    6.7 &    8.7 &   51.9 &    0.9 & 45.7  & 46.4  & CHI       & \ldots & \ldots &\\
 22611 A,B&   16.9 &    6.8 &    9.0 &   99.6 &    0.7 & 45.7  & 46.0  &  CHI      & \ldots  & \ldots & B=22604 \\
 22826 A,B&   19.9 &    7.2 &    8.7 &  415.9 &    0.3 & 12.1  & 12.0  & GCS      & \ldots & \ldots & B=HD 31222 \\
 23693 A,B&   85.9 &    4.8 &    9.0 &  321.7 &    0.6 & --0.8   &  --0.7  & GCS,CHI  & \ldots & \ldots & B=23708 \\
 23926 A,B&   18.7 &    6.8 &   10.3 &   10.1 &    1.9 & 43.8   & 45.1  & GCS,CHI   & \ldots & \ldots & B=23923 \\
 24711 A,B&   15.3 &    8.5 &   10.6 &   13.4 &    1.5 & --11.3  & --10.8 & GCS,CHI   & v & \ldots & B=24712 \\
 25082 A,B&   17.6 &    7.1 &   10.1 &    9.5 &    1.9 & 25.1   & 23.2  &  GCS,CHI  & \ldots & \ldots & \\
 27922 A,B&   42.4 &    7.6 &   10.6 &   10.6 &    2.3 & 42.1   & 44.0   & GCS,CHI  & \ldots & \ldots & \\
 31711 AB,C&  47.0 &    6.3 &    9.8 &  808.6 &    0.4 & 28.1  & 31.6   & GCS,CHI  & V,s,a & \ldots & C=31878 \\
 33705 A,B&   26.8 &    6.7 &    8.6 &  323.9 &    0.4 & 16.3   & 16.4   & GCS     & \ldots & \ldots & B=33691 \\
 34065 AB,C & 60.5 &    5.6 &    8.8 &  184.9 &    0.9 & 88.1   & S1   & GCS,S11   & C  & a,S1,v & C=34052 \\
 34065 A,B&   60.5 &    5.6 &    7.0 &   21.0 &    2.7 & 88.1   & 88.0 & GCS       & \ldots & \ldots & B=34069 \\
 36165 A,B&   31.2 &    7.1 &    8.1 &   17.7 &    1.8 & 66.7  & 66.3 & GCS      & s & s & B=36160 \\
 36485 AB,D&  22.0 &    7.4 &    8.0 &  112.0 &    0.8 & --1.0   &  S1   & GCS,H12  & C & S1,a,v & D=36497  \\
 36640 A,B&   35.8 &    6.0 &    8.7 &   23.4 &    1.7 & 54.8   & 56.3   & GCS,CHI & \ldots & \ldots & B=36642 \\
 37645 A,B&   19.6 &    7.1 &   10.4 &    9.6 &    2.3 & S1    & --16.4   & CfA,CHI & S1,v & \ldots & \\
 38908 A,BC&  61.7 &    5.6 &    9.9 &   60.6 &    1.3 & 14.7  & 17.0    & GCS,CHI  & \ldots & v & \\
 40452 A,B&   25.9 &    7.7 &    9.8 &   32.0 &    1.1 & 20.5  & 21.2    & GCS,CHI  & \ldots & \ldots & \\
 42488 A,B&   27.5 &    7.3 &    8.6 &   25.9 &    1.4 & --20.3  & --20.2  & GCS,C11 & s & \ldots & B=42491 \\
 44579 A,B&   15.5 &    8.7 &    9.3 &  140.5 &    0.4 &  2.2   & 0.6     & GCS,CHI   & \ldots & \ldots & B=HD 77759\\
 45734 A,B&   17.6 &    8.4 &    9.7 &    9.0 &    2.5 &  s2    & s2     &  CHI    & v,s & s2 & \\
 45838 C,AB&  16.5 &    7.5 &    9.2 &  135.0 &    0.6 &  52.6  & 55.0 & GCS,CHI     & \ldots & V        & AB=HD 60815\\
 46236 A,B&   21.4 &    7.0 &   10.2 &   19.3 &    1.4 &  30.2  & 31.0   & GCS,CHI     & \ldots & \ldots & \\
 47058 A,B&   21.3 &    7.9 &   10.5 &    8.1 &    2.1 &  60.0   &  S1   & GCS,CfA    & \ldots & S1      & \\
 47839 A,B&   19.0 &    8.1 &    8.2 &   18.8 &    1.8 &  --2.2   & --2.4  & GCS     & \ldots & \ldots & B=47836 \\
 47862 A,B&   15.8 &    7.2 &   10.8 &    9.6 &    1.8 &  --16.5  & --14.5  & GCS,CHI  & \ldots & \ldots &\\
 48785 A,B&   23.9 &    7.7 &    8.4 &   30.6 &    1.2 &  30.4  & 31.0    & GCS     & \ldots & \ldots & B=48786 \\
 49030 A,B&   16.7 &    7.7 &   10.5 &   24.2 &    1.1 & 14.7   & 15.8    &  GCS,CHI   & \ldots & \ldots &\\
 49520 A,B&   16.9 &    8.8 &    9.0 &    9.5 &    1.9 & --0.4    & 0.3     & D06    & \ldots & v        &\\
 56242 A,B&   42.9 &    6.4 &    9.2 &   15.5 &    2.1 & --4.9   & --4.9     & GCS   & \ldots & \ldots   & \\
 57148 A,BC&  16.0 &    8.4 &    9.8 &   24.9 &    1.2 & 4.3    & 4.3     &  LCO    & \ldots & v  & BC=57146 \\
 58067 A,B&   24.7 &    8.3 &    8.5 &   73.4 &    0.8 & 5.9  & 5.6     & GCS     & \ldots & \ldots & B=58703 \\
 58813 A,B&   18.3 &    8.1 &    8.8 &   23.2 &    1.2 & 6.0  & 5.5    &  GCS   & \ldots & \ldots    & B=58815 \\
 59272 A,B&   45.0 &    7.0 &    9.7 &    9.4 &    2.7 & 1.6   & 2.7    & GCS,CHI     & \ldots & \ldots & \\
 59690 A,B&   18.1 &    7.6 &    9.7 &   23.2 &    1.2 & 24.2  & 25.1   & GCS,CHI     & \ldots & \ldots & B=59687 \\
 60353 A,B&   33.4 &    6.6 &   10.5 &   20.7 &    1.7 & 4.3   & 4.4    & GCS,CHI     & \ldots & \ldots & B=60352 \\
 60749 AB,D&  15.9 &    8.5 &   10.2 &   24.7 &    1.3 & --3.9  & --1.8    & GCS,CHI    & V & \ldots & D=60750 \\
 61595 A,B&   17.8 &    8.3 &   10.9 &    8.4 &    1.9 & --13.7 &  --15.3  & GCS,CHI    & \ldots & \ldots & \\
 64478 A,B&   23.7 &    6.3 &    9.4 &   25.1 &    1.9 &  S2  &  S2     &  S90,CHI    & V+S2 & S2 &  \\
 64498 A,B&   18.0 &    7.6 &    9.9 &    9.3 &    2.2 & --12.0 & --15.7   & CHI        & v & \ldots & \\
 66121 A,B&   28.4 &    6.5 &    9.3 &   22.1 &    1.5 & --28.9 & --28.4   & GCS,CHI     & \ldots & \ldots & B=66125 \\
 66676 A,BC&  16.9 &    8.4 &    9.2 &   77.5 &    0.7 &  1.3 &  2.4    & CHI     & \ldots & v  & BC=HD 118735 \\
 67246 A,B&   31.6 &    6.4 &   10.2 &  488.5 &    0.3 &  --30.9& --30.4  & GCS,CHI     & \ldots & \ldots  & B=67291 \\
 67408 A,B&   33.9 &    6.7 &   10.2 &   11.6 &    2.2 &  3.2  & 2.9    & GCS,CHI     & \ldots & \ldots & \\
 69220 A,B&   20.8 &    8.4 &    9.7 &   57.5 &    0.8 &  48.9 & 50.0   & GCS,CHI     & \ldots & \ldots & B=69224 \\
 71682 A,CD&  23.4 &    7.1 &   10.0 &  123.0 &    0.7 &  11.1 & 13.2   & GCS,CHI     & a & v & CD=71686 \\
 72235 A,B&   24.0 &    8.6 &   10.8 &    9.1 &    2.1 &  9.1  & 9.1    & CHI     & \ldots & v & \\
 74930 A,B&   20.0 &    7.2 &    8.1 &   13.3 &    1.8 &  --35.9 & --36.7 & GCS     & \ldots & \ldots & B=74931\\
 74975 A,B&   39.4 &    5.1 &   10.1 &   11.4 &    2.6 &  54.2  & 55.1    & GCS,CHI     & \ldots & \ldots & \\
 75790 AB,C&  17.1 &    7.0 &   10.1 &    9.6 &    2.3 &  --12.2  &  --10.9 & LCO     & v & \ldots & \\
 76435 A,C &  20.6 &    9.1 &   10.6 &   13.5 &    1.8 &  5.4   &   7.3   & CHI     & \ldots & v & \\
 76888 A,B&   15.0 &    9.3 &    9.4 &   23.2 &    1.1 & 8.1    &  7.9    & CHI     & \ldots & \ldots & B=76891 \\
 78738 A,B&   39.6 &    7.5 &    8.1 &   11.8 &    2.3 & --31.8  &  --32.3  & GCS     & \ldots & \ldots & B=78739 \\
 79730 A,B&   21.1 &    7.2 &    8.7 &   19.8 &    1.4 &  --40.6 & --40.1   & GCS,CHI & \ldots & \ldots & \\
 83701 A,B&   17.1 &    7.9 &   10.0 &   97.2 &    0.6 &  --1.3  & --1.1    & GCS,CHI      & \ldots & \ldots & \\
 85342 A,B&   21.1 &    7.0 &    8.9 &  127.2 &    0.6 &  --15.9 & --16.6  & GCS     & \ldots & a,s,v & B=85326 \\
 90355 A,B&   27.3 &    7.9 &    8.4 &  608.4 &    0.3 &   S1   & --18.0   & H12     & A,S1 & \ldots & B=90365 \\
 91837 AB,C&  16.0 &    7.7 &   10.2 &   32.6 &    1.2 &  --4.2  & --4.1    & LCO     & V & \ldots   & \\
 93772 A,B&   15.2 &    7.0 &    8.9 &    9.4 &    1.9 &  --18.2 & --18.7   & GCS,TS02  & \ldots & \ldots & \\
 95106 A,B&   21.2 &    8.2 &   10.3 &   13.7 &    1.9 &  13.7 & s1     & GCS,FECH    & v,s & s & B=95110 \\
 95116 A,B&   15.9 &    7.1 &    9.3 &    8.8 &    2.0 &  --41.9  & --41.1   & CHI     & \ldots & \ldots & \\
 96979 A,B&   15.1 &    6.9 &    8.7 &   27.3 &    1.1 &  --41.9  & --41.6  & GCS     & \ldots & \ldots & B=96976 \\
 97508 AB,C&  17.8 &    7.4 &   10.4 &   13.9 &    1.9 &  8.3   &  8.0    &  CHI & V & \ldots  & \\
 99729 A,B&   16.4 &    7.8 &    8.1 &   43.3 &    0.9 &  --0.9  & --0.6    & N02     & \ldots & \ldots & B=99727 \\
101551 A,B&   16.7 &    7.9 &    9.6 &   10.4 &    1.7 &  21.4  &  21.4   & CHI     & \ldots & \ldots   & B=101549 \\
102418 A,CD&  15.3 &    8.8 &   10.5 &  539.2 &    0.3 &  10.7:  &  11.3   & SIM,CHI     & \ldots & v & \\
102655 A,B&   18.6 &    8.5 &    9.7 &  391.3 &    0.3 &  --2.6  &  --2.3   &  CHI    & \ldots & \ldots & B=HD 198016 \\
103311 AB,C & 21.9 &    7.4 &   10.7 &  324.9 &    0.4 &  --4.5  & --6.9    &  SACY    & v & \ldots & \\
105585 AB,C & 18.0 &    8.9 &    9.0 &   18.1 &    1.6 &  s2   &  3.6     &  CHI    & v,s2 & \ldots & C=105569 \\
105879 A,D &  15.8 &    7.3 &   10.0 &   44.0 &    0.9 &  s2   & 35.9     &  CHI    & s2,a & \ldots &  \\
106438 A,B&   20.4 &    7.8 &    9.8 &   40.3 &    0.9 &  --29.2 & --29.0   & CHI   & \ldots & \ldots & \\
106632 A,B&   16.5 &    8.8 &   10.0 &   11.4 &    1.6 &  16.7  &  16.3   & CHI     & \ldots & \ldots & B=106633 \\
110447 A,B&   15.0 &    9.1 &   10.2 &   12.6 &    1.4 &  1.4   &  2.5    & CHI     & \ldots & \ldots & B=110444 \\
110712 A,B&   43.4 &    6.1 &    8.9 &   20.6 &    2.1 &  14.6   & 15.3   & FECH   & \ldots & v & B=110719 \\
112201 A,B&   18.5 &    8.3 &    9.8 &  179.0 &    0.4 &  --13.3  & --12.9   &  GCS,CHI     & \ldots & \ldots & \\
113386 A,B&   16.3 &    7.6 &    9.2 &    9.1 &    1.9 &  0.5   &  0.9     & CHI     & \ldots & \ldots & \\
113579 A,B&   32.5 &    7.6 &    9.7 &  581.2 &    0.3 &  7.0   &  s3     & FECH     & \ldots & v & B=113597 \\
114378 A,B&   40.3 &    6.5 &   10.3 &   31.7 &    1.4 &  S1    & --1.4     &  Griff01    & S1 & \ldots & \\
114702 A,B&   25.6 &    8.1 &    8.6 &   24.9 &    1.6 &  --30.0 &  S1    &  GCS,T06    & \ldots & v+S1 & B=114703 \\
116063 A,B&   32.6 &    7.2 &    9.1 &   36.4 &    1.2 &  4.0   & 4.3    &  FECH    & \ldots & \ldots & \\
117391 A,B&   17.5 &    7.6 &   10.1 &  109.3 &    0.5 &  17.2  &  17.7   & GCS,CHI     & \ldots & \ldots & 
\enddata 
\tablerefs{
SIM: SIMBAD;
C11: \citet{Chubak2011};
CfA: D. Latham, 2012, private communication;
CHI: this work, CHIRON data;
D06: \citet{Desidera2006};
FECH: this work, FECH data;
GCS: \citet{N04};  
Griff01: \citet{Griffin2001};
H12: \citet{Halb2012};
L02: \citet{Latham2002};
LCO: \citet{LCO};
N02: \citet{Nid02};
S11: \citet{Sahlmann2011}; 
S90: \citet{Saar1990};
SACY: \citet{SACY};
SB9: \citet{SB9};
T06: \citet{Tok06};
TS02: \citet{TS02}.
}
\end{deluxetable}



\end{document}